\documentclass[twocolumn,tighten,times,floatfix]{aastex631}

\usepackage{amsmath}
\usepackage{anyfontsize}
\usepackage{amssymb}
\usepackage{sidecap}
\usepackage{natbib}
\usepackage{lipsum}
\usepackage{nameref}
\usepackage[T1]{fontenc}
\usepackage{graphicx}
\usepackage{hyperref}
\usepackage{svg}
\usepackage[normalem]{ulem}
\usepackage{soul}
\usepackage{placeins}
\usepackage{float}
\usepackage[flushleft]{threeparttable}
\usepackage{rotating}

\usepackage{newunicodechar}
\DeclareRobustCommand{\okina}{%
  \raisebox{\dimexpr\fontcharht\font`A-\height}{%
    \scalebox{0.8}{`}%
  }%
}
\newunicodechar{ʻ}{\okina}

\newcommand\blfootnote[1]{%
  \begingroup
  \renewcommand\thefootnote{}\footnote{#1}%
  \addtocounter{footnote}{-1}%
  \endgroup
}

\newcommand{\snana}{{\fontfamily{qcr}\selectfont{SNANA}}}

\newcommand{\saltshaker}{\texttt{SALTshaker}}

\def\one{{\,\sc i}}
\def\two{{\,\sc ii}}
\def\three{{\,\sc iii}}
\def\four{{\,\sc iv}}
\def\five{{\sc v}}
\def\six{{\sc vi}}

\defcitealias{kenworthy_salt3_2021}{K21}
\defcitealias{Pierel_salt3nir_2022}{P22}

\shorttitle{Feeling Blue: Extending SALT3 into the UV}
\shortauthors{Wang et al.}
\graphicspath{{./}{figures/}}

\begin{document}

\newcommand{\MIT}{Department of Physics and Kavli Institute for Astrophysics and Space Research, Massachusetts Institute of Technology, 77 Massachusetts Avenue, Cambridge, MA 02139, USA}
\newcommand{\JHU}{Physics and Astronomy Department, Johns Hopkins University, Baltimore, MD 21218, USA}
\newcommand{\STScI}{Space Telescope Science Institute, Baltimore, MD 21218, USA}
\newcommand{\UCSC}{Department of Astronomy and Astrophysics, University of California, Santa Cruz, CA 95064, USA}
\newcommand{\NSF}{National Science Foundation Graduate Research Fellow}
\newcommand{\UCAS}{School of Astronomy and Space Science, University of Chinese Academy of Sciences, Beijing 100049, People’s Republic of China}
\newcommand{\NAOC}{National Astronomical Observatories, Chinese Academy of Sciences, Beijing 100101, People’s Republic of China}
\newcommand{\StockholmU}{The Oskar Klein Centre, Department of Physics, Stockholm University, SE - 106 91 Stockholm, Sweden}
\newcommand{\UPitts}{Department of Physics and Astronomy and PITT PACC, University of Pittsburgh, Pittsburgh, PA 15260, USA}
\newcommand{\UHIfA}{Institute for Astronomy, University of Hawai’i, 640 N. A’ohoku Pl., Hilo, HI 96720, USA}
\newcommand{\UMD}{Department of Astronomy, University of Maryland, College Park, MD, 20742-2421, USA}
\newcommand{\KICP}{Kavli Institute for Cosmological Physics, University of Chicago, Chicago, IL 60637, USA}
\newcommand{\UChicago}{Department of Astronomy and Astrophysics, University of Chicago, Chicago, IL 60637, USA}
\newcommand{\UT}{Department of Astronomy, The University of Texas at Austin, 2515 Speedway, Stop C1400, Austin, TX 78712, USA}
\newcommand{\SSI}{Space Science Institute, Lawrence Livermore National Laboratory, 7000 East Avenue, Livermore, CA 94550, USA}

\title{Feeling Blue: Constructing a Robust SALT3 UV Template and Constraining its Redshift Dependency}

\author[0000-0001-5233-6989]{Qinan~Wang}\blfootnote{Corresponding author: Qinan~Wang\\ \href{mailto:qnwang@mit.edu}{qnwang@mit.edu}}
\affiliation{\MIT}
\author[0000-0002-6230-0151]{David~O.~Jones}
\affiliation{\UHIfA}
\author[0000-0002-2361-7201]{Justin~D.~R.~Pierel}
\affiliation{\STScI}
\affiliation{NASA Einstein Fellow}
\author[0000-0003-2445-3891]{Matthew~R.~Siebert}
\affiliation{\STScI}
\author[0000-0002-5153-5983]{W.~D'Arcy~Kenworthy}
\affiliation{\StockholmU}
\author[0000-0003-3221-0419]{Richard Kessler}
\affiliation{\KICP}
\affiliation{\UChicago}
\author[0000-0002-5995-9692]{Mi~Dai} 
\affiliation{\UPitts}
\author[0000-0002-2445-5275]{Ryan~J.~Foley}
\affiliation{\UCSC}
\author[0000-0003-2238-1572]{Ori~D.~Fox}
\affiliation{\STScI}
\author[0000-0003-3703-5154]{Suvi~Gezari}
\affiliation{\UMD}
\affiliation{\JHU}
\author[0000-0001-6395-6702]{Sebastian~Gomez}
\affiliation{\UT}
\author[0000-0002-1052-6749]{Peter~McGill}
\affiliation{\UCSC}
\affiliation{\SSI}
\author[0000-0002-4410-5387]{Armin~Rest}
\affiliation{\STScI}
\affiliation{\JHU}
\author[0000-0002-7559-315X]{César~Rojas-Bravo}
\affiliation{\UCAS}
\affiliation{\NAOC}
\author[0000-0002-9301-5302]{Melissa~Shahbandeh}
\affiliation{\STScI}
\author[0000-0002-7756-4440]{Lou~Strolger}
\affiliation{\STScI}


\begin{abstract}

Upcoming cosmological surveys will obtain numerous rest-frame ultraviolet (UV) observations of Type Ia supernovae (SNe~Ia), yet there is concern about how standardizable SNe~Ia are in the UV. 
In this work, we train a robust optical--UV SED model for SNe~Ia (SALT3-UV) with the open-source model-training software \saltshaker. We incorporate a spectroscopic UV data sample from HST, including 67 UV spectra from 18 nearby SNe~Ia. Unlike previous training spectra, the HST spectra have sufficiently precise calibration that they do not require additional warping to match coincident photometric data.  Additionally, while including this new SN~Ia sample necessitates incorporating auxiliary photometric data from ZTF and ATLAS that have insufficient calibration for cosmological analyses, the improvements in the calibration of these data is anticipated in the near future.
Compared to the previous SALT3-K21 model, the SALT3-UV model shows a significant improvement  in the UV down to $2000$\AA, with over a threefold improvement in model uncertainty and a more physically accurate continuum and line features. We further evaluate potential redshift evolution in the UV template by separating the UV training sample into low- and high-$z$ subsamples. Our results reveal a non-negligible $\gtrsim 0.05$ mag difference between low- and high-$z$ SALT3-UV models in the $g-$band at $z\gtrsim0.5$ and the $u-$band at $z\gtrsim0.2$. We demonstrate that, if confirmed, such evolution could lead to a few-percent bias in the measurement of $w$ if high-$z$ rest-frame UV data are included in future cosmological surveys such as LSST and \textit{Roman}.

\end{abstract}



\section{Introduction} \label{sec:intro}

In the next decade, cosmological surveys will leverage the enormous amount of Type Ia Supernovae (SNe~Ia) discovered from deep surveys, including the Vera C.\ Rubin Observatory's Legacy Survey of Space and Time (LSST) and the \textit{Nancy Grace Roman Space Telescope}, from the local universe up to $z\sim 3$ \citep{Rose21, Mitra23, Kessler25, Rubin25}. With rest-frame ultraviolet (UV) wavelengths becoming redshifted into optical observer-frame bands at high redshifts, the reliability of SN~Ia distances in the UV will become more and more important for measurements of the dark energy equation-of-state parameter, $w$, particularly for LSST \citep{lsst19}.

Recent measurements of $w$ use increasingly large SN\,Ia sample sizes and have become increasingly close to being limited by systematic uncertainties \citep{Scolnic18,Jones19,Brout22,Abbott24,Vincenzi24}. These measurements rely on the SALT2 \citep{Guy07,Guy10,Betoule14,Taylor21} or SALT3 (\citealt{kenworthy_salt3_2021}, hereafter SALT3-K21; \citealt{Pierel_salt3nir_2022,Taylor23}) models, which are spectrophotometric models to standardize SNe and measure their distances.  However, these models are not well-trained in the UV ($\lesssim 3500$\AA) due to the scarcity of well-calibrated spectroscopic and photometric UV data \citep{Taylor23}. The previous SALT3 training sample is comprised of just 85 SN~Ia spectra that cover UV wavelengths, with very limited coverage in phase and wavelength range and a median S/N of just $\sim$1.4.
The majority of these spectra are at $z>0.2$, introducing concerns about whether the redshift-evolving spectra of SNe\,Ia could bias $w$ \citep{milne13, foley13, foley16, Milne2015, Pan20, Nicolas21, Thorp24, Popovic25b}.

At high redshifts, where we obtain the maximum lever arm to constrain $w$, the UV SED model becomes more and more important. For example, a robust model below $3500$\AA\ will be necessary to fit $\sim35\%$ of all \textit{Roman} SNe~Ia \citep{Rose21}. LSST will also have extensive $u$-band coverage with enough sensitivity to detect SNe~Ia out to redshifts $z \simeq 0.3$. Meanwhile, JWST has started to probe and build a sample of SNe~Ia at $z\gtrsim2$ and extended the distance ladder into the dark matter dominated era \citep[e.g.][]{Pierel24, snencore, 2025ogs}. However, due to the large uncertainties in the UV templates and concerns about its redshift dependence, recent surveys tend to exclude those rest-frame UV data. For example, the Dark Energy Survey’s five-year SN survey has excluded data with $\lambda_{\textrm{rest-frame}}<3500$\AA\ in its analysis \citep{Abbott19}.

Modeling UV wavelengths can also help to constrain the SN\,Ia color law, a combination of intrinsic variation and dust attenuation that correlates with luminosity.  While the SALT3 color law closely follows dust attenuation laws \citep[e.g.,][]{Fitzpatrick99}, redward of  $\sim 4000$\AA\ it strongly diverges in the blue \citep{kenworthy_salt3_2021}.  The SALT color law --- and the role of dust attenuation laws in SN~Ia distances as a whole --- are responsible for some of the largest systematic uncertainties in cosmology today \citep{Brout21,Brout22,Popovic23}. 
The UV wavelengths are the most sensitive to dust attenuation and how it may vary as a function of redshift. \cite{Dai23} specifically tested how SALT3 model surfaces and distances changed as a function of the training sample, and found that the observer frame $u$/$U$-band data --- despite their often poor calibration --- can be important for constraining the SALT3 color law and model surfaces.

Also potentially problematic for high-redshift SN\,Ia distance measurements are a number of studies showing that SNe\,Ia have significantly larger diversity in the UV than in the optical \citep{Wang+12, milne13, foley13, foley16, Milne2015, Pan20, Nicolas21, Hoogendam24}. \cite{Foley08b} find that the SED template of SNe~Ia shows an intrinsic variation of $>10\%$ in the rest-frame UV, significantly greater than the $3\%$ variation they find in the optical region, putting doubt on the reliability of SNe~Ia as standardizable candles at shorter wavelengths. 
This diversity may stem from the fact that UV SED of SNe\,Ia are particularly sensitive to progenitor properties such as metallicity, mass, temperature, ionization and dust \citep{Hoflich98, Lentz00, Sauer08, Hachinger13, Mazzali14, Polin19}. Observational evidence also shows that UV spectra of SNe~Ia are correlated with the metallicity of their host galaxies, with UV-bright SNe~Ia tending to reside in metal-poor host galaxies \citep[][]{Pan20}.

The dependence of the SN~Ia UV SED on its progenitor or host-galaxy properties could imply the redshift evolution of the mean UV spectrum of SNe\,Ia. For example, with lower metallicity at high redshift, one might expect weaker line blanketing from iron-group elements and thus more UV-luminous SNe\,Ia.
Another possibility is that if there exists a wide range of binaries with diverse properties or even multiple progenitor channels contributing to the normal SNe~Ia population, the evolution of their properties and relative rate with cosmic time could lead to redshift dependence \citep[e.g.,][]{Rigault13,Childress14}.
Highlighting how UV variation may introduce redshift-dependent systematics in cosmological parameter measurements,
\cite{Milne2015} found that roughly two-thirds of low-$z$ SNe\,Ia are in a ``UV-red'' group, while ``UV-blue'' SNe\,Ia dominate in the intermediate and high redshift range.
Such redshift dependence, if it exists, would challenge the use of SNe~Ia as standardizable candles in the rest-frame UV regime. On the other hand, if the same physical mechanism introduces subtle bias in optical bands such as the $B-$ and $g-$band, the UV properties could provide a crucial diagnostic to mitigate it.
Assessing the impact and utility of UV SN\,Ia observations for cosmological parameter measurements is therefore crucial for deciding whether rest-frame UV data could be usable in upcoming cosmological surveys.

Training a robust SN\,Ia model in the UV has historically been difficult due to the challenges of precisely calibrating UV photometry from the ground.  At $z \gtrsim 0.3$, the $g$ band begins to probe the rest-frame UV, but at lower redshifts few ongoing time-domain surveys have well-calibrated measurements at UV wavelengths. However, numerous {\it Hubble Space Telescope} ({\it HST}) programs over the last 25 years have observed UV spectra of SNe\,Ia with the STIS instrument, which is capable of extremely well-calibrated spectrophotometric observations at UV wavelengths \citep{Bohlin20, Bohlin20b}.
Whereas SALT must typically recalibrate its spectral data during training, which can otherwise introduce systematic errors in the model \citep{Dai23}, this step is unnecessary for the already well-calibrated STIS data
Here, we use this \textit{HST}/STIS UV spectra sample to train a new SALT3 model with improved UV data sample (hereafter SALT3-UV).  The new dataset doubles the existing UV training sample for SALT3 with dramatic improvement in S/N.  We use this model to explore the ways in which SN~Ia SEDs may be systematically changing as a function of redshift, and the resulting implications for cosmological parameter measurements.
While we currently pair some of UV data with optical data that have 1-2\% systematic calibration uncertainties from ZTF and ATLAS, as those data are updated in the future, a re-training can be easily done with SALT3 framework to provide a precise template for cosmology measurements. 

In Section \ref{sec:data} we review the existing data and discuss the updated SALT3 training sample.  In Section \ref{sec:training} we review the \saltshaker\ framework and present the training. In Section \ref{sec:validation} we discuss validation of our trained model with cosmological simulation. In Section \ref{sec:discussion} we discuss the implications of our results for cosmology and SN\,Ia physics, and in Section \ref{sec:conclusion} we conclude.

\section{SALT3-UV Training Data} \label{sec:data}

Here, we discuss our updates to the SALT3-UV training sample using additional UV spectroscopy from HST and complimentary photometry for those same SNe.  The data comprising our sample are described below.

\subsection{Previous SALT3 Training Data}

The previous SALT3 training data, described in \citet{kenworthy_salt3_2021}, is a compilation of more than 30~years of combined SN\,Ia data from multiple surveys.  It builds on previous training data sets for SALT2 and prior models.  In brief, these include low-redshift SNe from the CfA samples \citep{Riess99, Jha06, Hicken12}, the Carnegie Supernova Project's Second Data Release \citep{Stritzinger11}, the Foundation Supernova Survey \citep{Foley18,Jones19}, as well as higher-redshift samples from the Sloan Digital Sky Survey \citep{Sako18}, the Pan-STARRS survey \citep{Scolnic18}, the Dark Energy Survey's three-year spectroscopic analysis \citep{Abbott19}, and the Supernova Legacy Survey \citep{Astier06}.  In total, the training sample includes 1083 SNe\,Ia with 1207 spectra.

These data were calibrated with the ``Supercal'' cross-calibration procedure, which used the Pan-STARRS 3$\pi$ coverage to match all surveys to a common calibration framework.  This was subsequently updated with the ``Fragilistic'' calibration for the Pantheon$+$ data release \citep{Brout22,Brout21, Taylor23}, which also updated the Pantheon CSP photometry from the second to the third CSP data release \citep{Krisciunas17}.

In this study we re-examine the rest-frame UV data from K21, and remove several non-{\it HST} rest-frame UV spectra without contemporary photometry for calibration, including SN~2005cf, SN~1992A, SN~1998B, SN~1990N and SN~1991T.  
\cite{Dai23} found that the lack of well-calibrated photometry at UV wavelengths for SNe with UV spectral coverage could introduce significant systematic uncertainties. Therefore, we conservatively opt to remove these spectra for the present analysis to help constrain the best-fit model parameters.  

\begin{figure}
    \centering
    \includegraphics[width = 0.99\linewidth]{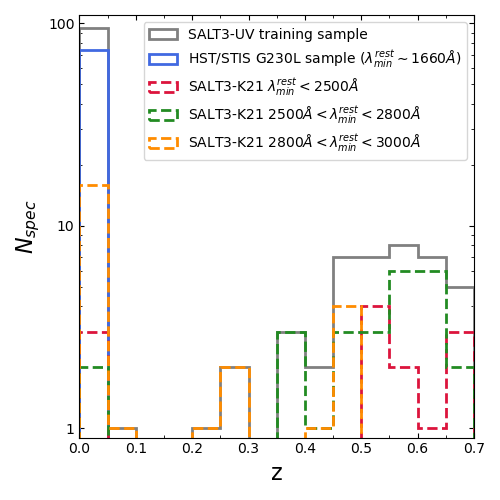}
    \caption{The number of UV spectra in the training sample as a function of redshift. The blue solid line represents the new \textit{HST}/STIS G230L spectra in the SALT3-UV training sample, all of which are at low $z$ and with minimum wavelengths around $1660$\AA. Dashed lines in different colors represent the spectra in SALT3-K21 with different minimum wavelengths in the rest frame, $\lambda_{min}^{rest}$, and the black solid line shows the summed training sample for the SALT3-UV model.  Most low-$z$ UV spectra in the previous SALT3-K21 training sample only marginally probe the UV, and at $\lambda<2500$\AA\ the SALT3-K21 sample is dominated by low S/N spectra at high $z$. }
    \label{fig:nspec-z}
\end{figure}

Due to the relatively low throughput of existing UV instruments, the rest-frame spectroscopic UV data, in particular at $\lambda_{\textrm{rest-frame}}<2500$\AA, are collected from either UV instruments observing nearby SNe ($z\lesssim0.02$) or optical instruments observing more distant SNe ($z\gtrsim0.2$). Therefore, we set $z = 0.1$ as the threshold for the low- and high-$z$ subsamples in the later discussions, not only because of this natural data gap, but also to make the data density comparable between the two subsamples.
The original SALT3-K21 sample includes approximately 300~SNe\,Ia at $z < 0.1$, with the remaining $\sim$800~SNe\,Ia at higher redshift. On the other hand, 1132 of the spectra in SALT3-K21 sample are from these low-$z$ samples, with only 63 and 2 high-$z$ spectra originating from the SNLS and Foundation samples, respectively. 
The K21 sample contains $64$ spectra extended into the rest-frame UV, including 22 spectra in the low-$z$ subsample ($z<0.1$) and 42 in the high-$z$ subsample. However, the majority of those spectra in the low-$z$ subsample do not sufficiently cover the UV wavelength range. In particular, as shown in Figure~\ref{fig:nspec-z}, among the low-$z$ subsample, only 5 spectra from SN~1994ae and SN~2001ep extend below $2800$\AA, while the high-$z$ subsample contains $34$ spectra in the same wavelength range. Thus, the UV part of the SALT3-K21 model is mainly based on the high-$z$ sample, which has relatively low S/N.  Due to this deficit of data, the K21 model is significantly smoothed by regularization, which penalizes large model variations in regions with few data points.

\begin{table}
\caption{\label{tab:1} Details of the UV spectroscopic data in the SALT3-UV training sample. All the new HST spectroscopic data come from SNe~Ia at low-$z$ with $z<0.1$. Despite the considerable number of spectra in the SALT3-K21 sample, the majority of SALT3-K21 spectra do not sufficiently cover the UV wavelength range between $2000-3000$\AA\ (see discussion in Section \ref{sec:data}). 
}
\begin{tabular}{c|c|c}
\hline
\hline
 Survey & $N_{\rm SN}$ &  No. of UV Spectra$^a$\\
\hline
{\it HST}/STIS &18 &67 \\
\hline
K21 low-$z$$^b$&12&22\\
\hline
K21 high-$z$$^c$&36&42\\
\hline
\hline
low-$z$&29&87$^d$\\
high-$z$&36&42\\
\hline
\hline
\textbf{UV Training Sample Total}&\textbf{65}&\textbf{129}\\
\hline
\hline
\end{tabular}
$^a$Includes spectra with rest-frame wavelength coverage $\lambda\leq 3000$\AA.\\
$^b$$z<0.1$\\
$^c$$z>0.1$\\
$^d$Two of the SN~2001ep {\it HST} spectra have duplicates in K21 low-$z$ sample.\\

\end{table}

\subsection{HST UV data}

To create the SALT3-UV training set, we search for SNe~Ia that have combined STIS (G230L, G430L and G750L) spectra in {\it HST} archive and corresponding optical light curves sufficient for measuring light curve parameters. 
The G230LB spectra are not included due to the systematic calibration differences in the redder sources caused by the scattered light \citep{Worthey22}.
In order to accurately constrain the epoch of peak brightness, we require that the light curve begin at or before peak, and extend to at least $+15$~days after peak. Each spectrum must have a rest-frame phase in the SALT3 model range to be included ($-20$ to $+50$ days relative to peak).
The optical light-curve data used to determine the time of maximum light are described in Section~\ref{sec:photo}.

In the end, the {\it HST}/STIS sample contains 67 spectra from 18 bright, nearby (median $z \sim 0.005$) SNe~Ia taken by {\it HST} programs GO-9114 (PI R.P.Kirshner), 11721, 12298 (PI R.Ellis), 12582 (PI A.Goobar), 13286, 13646, 14925, 16238, 16690 (PI R.Foley), 14665, 16190, 16221 (PI P.Brown), and 17170 (PI M.Siebert). All the {\it HST} data used in this paper can be found in MAST: \dataset[10.17909/kr7g-7z02]{http://dx.doi.org/10.17909/kr7g-7z02}.
Table \ref{tab:1} summarizes the statistical information of SNe~Ia and UV spectra used in this training, including this new sample that were not used in the previous SALT3-K21 model training, and Figure~\ref{fig:nspec-z} highlights the difference in the redshift distribution and wavelength coverage between the previous and new spectroscopic samples. 
Details of those new SNe are listed in Table~\ref{tab:sninfo}. These spectra provide continuous, high-precision, well-calibrated rest-frame UV information from $1700-5500$\AA, in particular below $2500$\AA\ where only 13 spectra in SALT3-K21 sample provided coverage. There are two reasons that those \textit{HST} spectra were not utilized in previous trainings: firstly, many of the SNe in our new sample are not contained in previous training samples because they were not observed as part of the surveys like CfA or CSP; secondly, these \textit{HST}/STIS spectra were not uniformly calibrated until recently.

We processed the STIS data with the ``calstis'' spectroscopic pipeline (part of the python package \texttt{stistools}\footnote{\url{https://stistools.readthedocs.io/en/latest/}}, ver. 1.4.4) to bias subtract, flat-field, extract, wavelength-calibrate, and flux-calibrate each SN spectrum. For visits that contained G230L, G430L, and G750L, the data have been combined in their overlapping regions (weighted by S/N). STIS MAMA spectra can reach an absolute photometric accuracy of $4\%$ --- which does not affect the SALT training in this work, as we marginalize over the absolute spectral flux --- and a relative accuracy of $2\%$ alone \citep{stishandbook}, sufficient for training a reliable SNe~Ia spectrophotometric template in UV.


\begin{figure*}
    \centering
    \includegraphics[width = 0.98\linewidth]{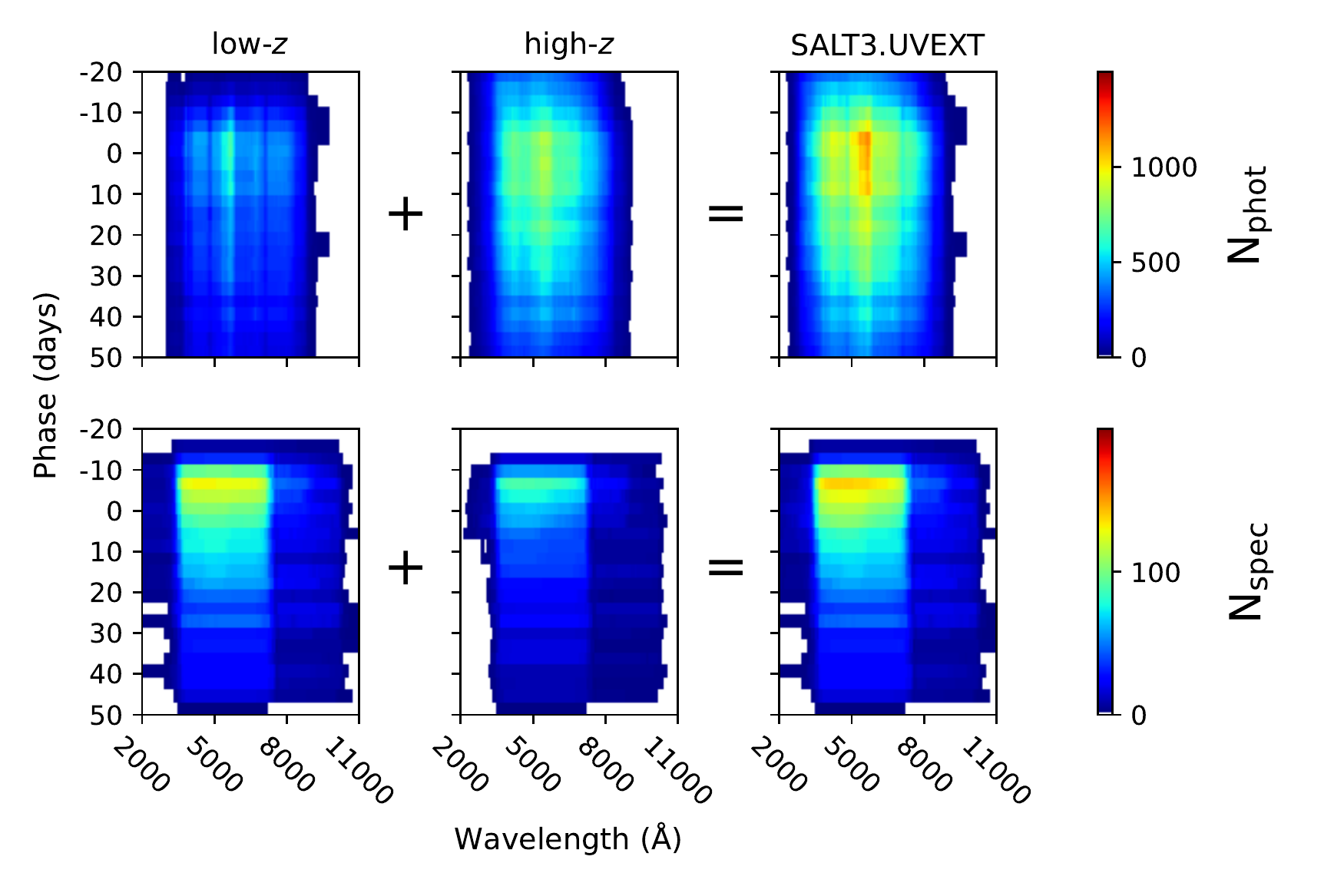}
    \caption{The density of photometric (top) and spectroscopic (bottom) data in our SALT3-UV training sample, as a function of rest-frame wavelength and days relative to peak brightness (phase). The sample can be divided into low-$z$ ($z<0.1$, left) and high-$z$ ($z\geq0.1$, middle) subsamples with distinct UV data. The low-$z$ subsample and high-$z$ subsample (supplemented with low-$z$ optical data) are used for separate model trainings in Section~\ref{sec:ztraining} to evaluate potential redshift dependence in the UV. }
    \label{fig:data-density}
\end{figure*}

\subsection{Optical and UV Photometry}\label{sec:photo}

For each SN with {\it HST} UV spectra, we compile the available photometry from the Pantheon$+$ analysis \citep{Scolnic22,Brout22}, the existing SALT3-K21 training sample \citep{kenworthy_salt3_2021}, the Asteroid Terrestrial-impact Last Alert System \citep[ATLAS;][]{Tonry18}, and the Zwicky Transient Facility \citep[ZTF;][]{Bellm19}, with details listed in Table~\ref{tab:sninfo}. Although the ATLAS and ZTF photometry may have higher than typical calibration uncertainties as discussed below, these photometry are still necessary; the primary reason is that cadenced optical data allow precise measurements of the SALT $x_0$, $x_1$, and $c$ parameters for each SN so that the training is able to then model the UV SED as a function of those parameters. We note that these photometric data only constitute a relatively small portion of the complete training sample (five of the SNe in the training sample include data from ZTF and another five include ATLAS data), and SALT3 can be easily re-trained after subsequent ATLAS and ZTF calibration improvements.  Below, we briefly describe the ZTF and ATLAS SN data.

ZTF is an optical time-domain survey that observes the entire visible sky at a two-day cadence in the $gri$ filters \citep{Bellm19}. ZTF's SN\,Ia data release papers note that the photometry is not yet accurate enough for competitive cosmological parameter inference, which will be provided in a subsequent Data Release 2.5 \citep{Rigault25}. \citet{Kenworthy25} also points to the existence of unmodelled (potentially calibration-induced) uncertainties at the 1-2\% level in the ZTF data. Recent calibration efforts also revealed that some non-linearity effects in their photometry that may lead to additional bias \citep{Lacroix25}. 

ATLAS is an all-sky time-domain survey with telescopes at four sites: Hawaiʻi (Haleakala and Maunaloa), Chile, and South Africa.  ATLAS observes the visible sky multiple times per night in the ``cyan'' and ``orange'' bands, approximately equal to wide-band $g+r$ and $r+i$ filters.  Although ATLAS claims a 5~mmag calibration uncertainty, this has not yet been fully validated; \citet{Scolnic25} notes that there is a potential color-dependent calibration offset affecting the cyan band at the 0.04~mag level. Recent work by \cite{Marlin25} has cross-calibrated the ATLAS forced photometry from the multiple sites to the Dark Energy Survey (DES) YR6 release, reaching combined calibration-related systematics of $\sim 5-10$mmag. The full cosmology-grade light curve sample will be released as The Type Ia supernova Trove from ATLAS in the Nearby universe (TITAN) DR1 soon (Murakami et al. in prep, Tweddle et al. in prep).

The ZTF and ATLAS data used here are from their respective forced-photometry servers, described in \citet{Masci19} and at the \href{fallingstar.com}{ATLAS website}, respectively. Both perform PSF-fitting photometry at fixed locations on the sky provided by the user. {\it Swift} data in our sample come from the Pantheon$+$ compilation \citep{Scolnic22} and are recalibrated by the Fragilistic method \citep{Brout21}.  The data were originally retrieved from the Swift's Optical/Ultraviolet Supernova Archive \citep[{\tt SOUSA};][]{Brown14}. We use only the $UBV$ bands in the training as the original calibration in the UVW1, UVM2 and UVW2 bands comes with $\delta zp \sim0.03$ mag and thus does not satisfy the precision level for cosmological analysis \citep{Brown09, Brown10, Brown14}. Re-calibration of the {\it Swift} UV data is nearly impossible as these bands are not covered by other surveys.

\section{SALT3 Model Training} \label{sec:training}

We use the SALT3 model training pipeline, named \saltshaker\ \citep{kenworthy_salt3_2021}, to simultaneously fit the parameters of the SALT3 model to our training data.  The SALT3 training process is described extensively in Section 2 of K21 paper, and we describe it briefly below.

SALT3 follows previous versions of SALT in using a principal component plus color-law model to describe the spectral surfaces of SNe~Ia.  The phase- and wavelength-dependent flux, $F(p,\lambda)$, of a given SN~Ia is described by:

\begin{align}
        F(p,\lambda) = &x_0 [M_0(p,\lambda;\boldsymbol{m_0}) + x_1 M_1(p,\lambda;\boldsymbol{m_1})] \nonumber\\ & \cdot \exp(c \cdot CL(\lambda;\boldsymbol{cl})).    
\end{align}

\noindent The SALT model flux is defined by the phase- and wavelength-dependent zeroth and first principal component surfaces, $M_0$ and $M_1$, which depend on a cubic spline interpolation with knots ${\bf m_0}$ and ${\bf m_1}$, and the color law.  The color law is a flexible polynomial parameterization with coefficients ${\bf cl}$.  These components are scaled by the individual parameters for each SN: the amplitude, $x_0$, the scale of the first principal component ($x_1$; often called the shape parameter as it correlates strongly with light-curve stretch), and the color parameter, $c$.

SALT3's model definitions place a number of constraints on the model surfaces to avoid degeneracies. The amplitude, $x_0$, is a free parameter, which allows SALT3 to be cosmology independent; for this reason, the magnitude of $M_0$ in the $B$-band at maximum light is fixed to be $m_\textrm{B}^\textrm{peak}=10.5$ when $x_0=1$.  To remove the degeneracy between the amplitude of $M_1$ and the values of $x_1$, the mean $x_1$ and it's standard deviation are defined to be (0,1). The color-law is defined with $CL(4300$\AA$) = 0$ and $CL(5430$\AA$) = -1$, corresponding to the central wavelengths of the $B$ and $V$ bands respectively.

Together, these surfaces are determined by simultaneous fitting to more than 10,000 free parameters.  The model is compared to the data using the $\chi^2$, which is minimized via a gradient descent method implemented using {\tt jax} \citep{jax2018github}.  Errors are estimated iteratively with a log-likelihood approach.  This procedure is described in \citet{kenworthy_salt3_2021} with several updates in \citet{Kenworthy25}.

Because the model has more parameters than can be constrained by the data in some wavelength and phase regimes, \saltshaker\ adopts a regularization scheme that penalizes the $\chi^2$ for large variations as a function of wavelength and phase (``gradient'' regularization).  It also penalizes correlations between the phase- and wavelength-dependent model (``dyadic'' regularization).  The parameters scaling the amplitude of the regularization must be adjusted for each model training as the optimal regularization strength depends sensitively on the data density; tuning these parameters avoids ringing noise or over-smoothing of the model surfaces.

\subsection{{\tt SALTShaker} Parameters for UV Training}

In this analysis, we use a slightly extended wavelength range compared to previous SALT model trainings, with the model defined between $1800$ and $11000$\AA\ and the effective wavelengths of photometric filters restricted to be between $2000$ and $8700$\AA.  These restrictions omit portions of our spectra that extend below $1800$\AA, but little data exist to train this portion of the model. We expand the range of color law polynomials from 2800--8000\AA\ in SALT3-K21 to 2000--8000\AA, and the color law is linear on either side of this range.

We apply two primary modifications to the training process. 
First, we lower the regularization strength, particularly the phase and wavelength-dependent regularization and the regularization applied to the $M_1$ surface, finding that these are over-smoothed in our training especially in the UV; the gradients of spectra are naturally more abrupt at shorter wavelengths. Second, we update the \saltshaker\ training code to allow the {\it HST} spectra to {\it not} be recalibrated; the primary advantage of using STIS spectra to define our model is that they are extremely well calibrated already and do not require additional mangling to match the photometrically determined model surface \citep{Bohlin20, Bohlin20b}.

\subsection{Model Training Results and Comparison with SALT3-K21}

\begin{figure}
    \centering
    \includegraphics[width = 0.99\linewidth]{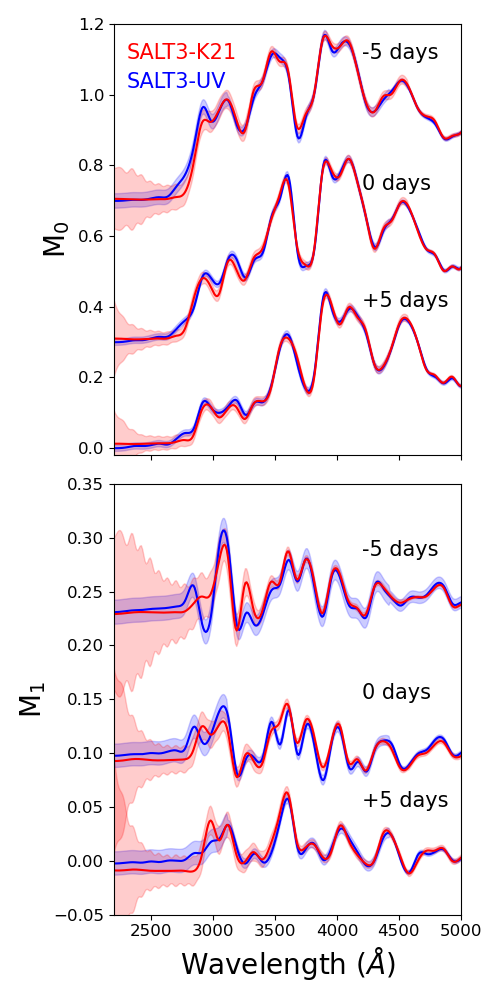}
    \caption{The comparison between the $M_0$ and $M_1$ component of the SALT3-UV and SALT3-K21 model between $2200-5000$\AA
    . Note that the relatively high regularization factors of the SALT3-K21 model over-smooth the UV templates and create a non-vanishing continuum at the short wavelength end. 
    }
    \label{fig:SALTmodelcomp}
\end{figure}

\begin{figure*}
    \centering
    \includegraphics[width=0.98\linewidth]{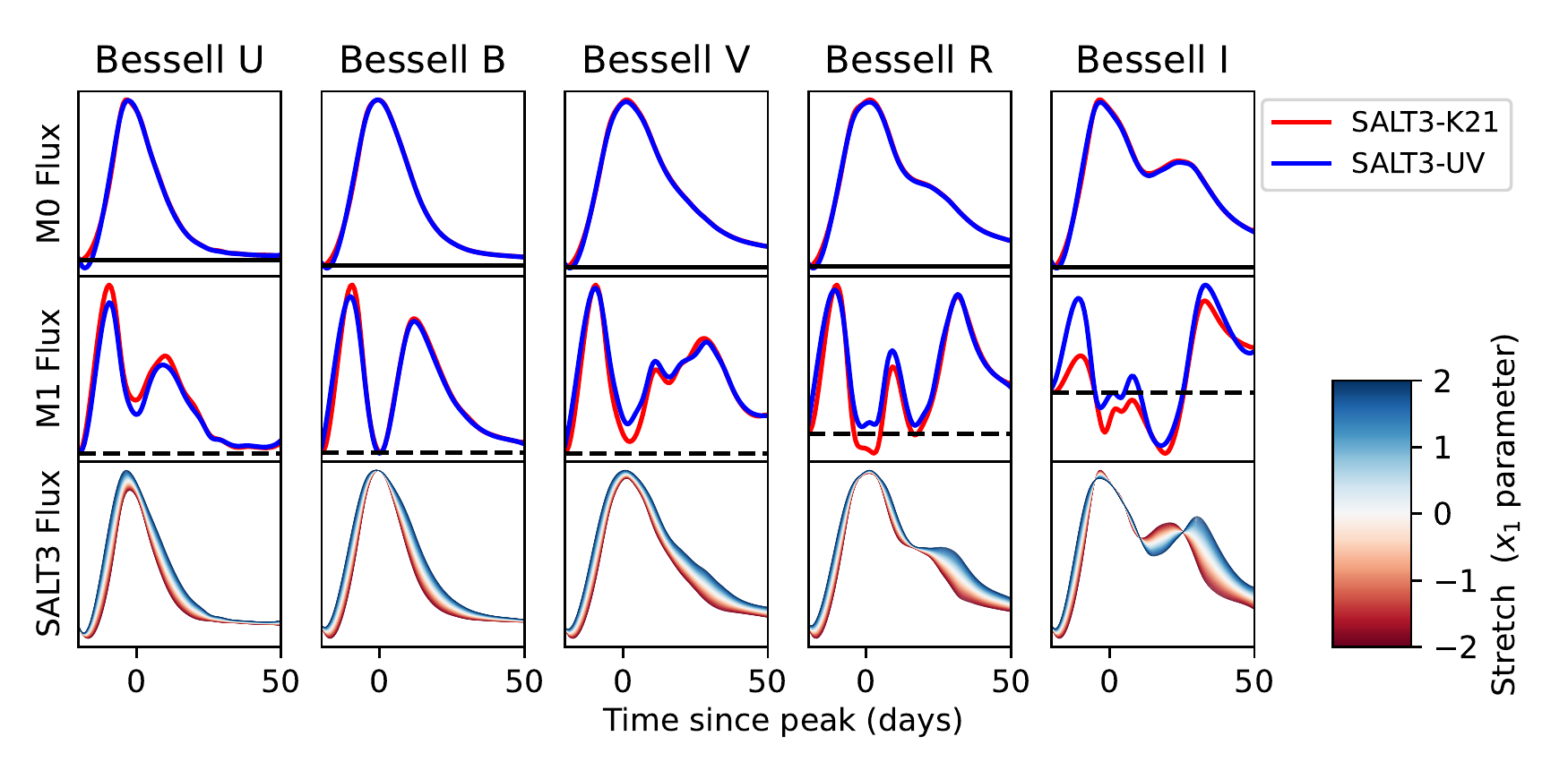}
    \caption{The SALT3-UV model flux integrated over optical bands in comparison with the SALT3.K21 model, as a function of phase and $x_1$ parameter. 
    }
    \label{fig:photooverplot}
\end{figure*}

\begin{figure*}
    \centering    \includegraphics[width=0.49\linewidth]{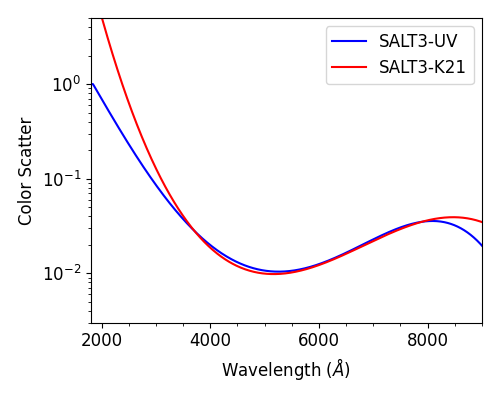}
\includegraphics[width=0.49\linewidth]{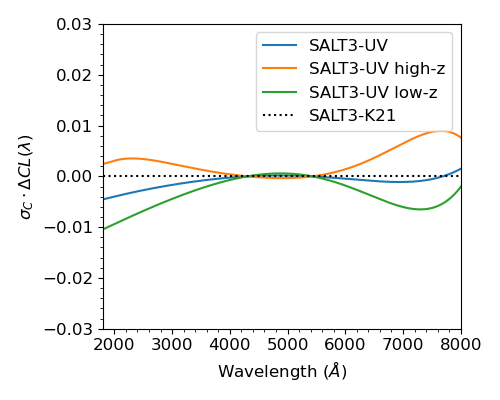}
    \caption{\textbf{Left:} Comparison between the color scatter of the SALT3-K21 and SALT3-UV models. \textbf{Right:} Difference between the color law of the SALT3-UV model and the SALT3-K21 model. $\sigma_c = 0.1$ is the standard deviation of the color parameter $c$ in a typical SN~Ia sample.}
    \label{fig:colorscatter}
\end{figure*}

\begin{figure}
    \centering
    \includegraphics[width=\linewidth]{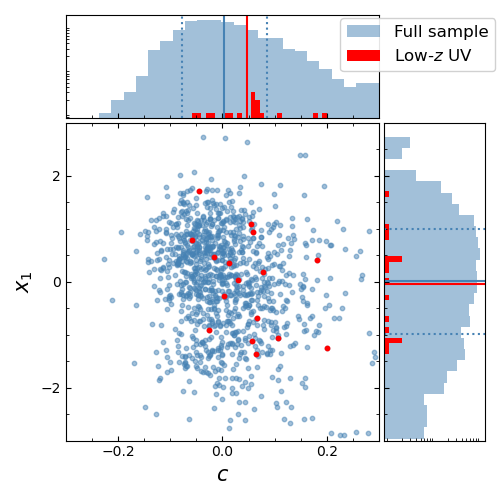}
    \caption{Distribution of $x_1$ and $c$ of the training sample as measured by SALT3-UV model. The newly added SNe~Ia with HST UV spectra are highlighted in red. The histograms are shown in logarithmic scale. The means and the standard deviation range of full sample are shown as solid and dotted lines in each histogram. }
    \label{fig:paramdist}
\end{figure}

Following the methods described above, we train the SALT3-UV model. The density of data as a function of phase and wavelength are shown in Figure \ref{fig:data-density}, divided into low-$z$ ($z\leq0.1$) and high-$z$ ($z>0.1$) subsamples. The model surfaces $M_0$ and $M_1$ are shown in Figure \ref{fig:SALTmodelcomp}, and corresponding model light curves in $UBVRI$ bands are shown in \ref{fig:photooverplot} in comparison with SALT3-K21 model. 
The `color scatter' term, defined as the relative covariance of the SALT model flux at two wavelengths, is modeled by a fourth order polynomial \citep{Guy10, kenworthy_salt3_2021};
the comparison between the color scatter of the previous fragilistic SALT3-K21 model versus SALT3-UV is shown in Figure \ref{fig:colorscatter}.  This shows a significant improvement in the color scatter below $3000$\AA\ by a factor of $\sim 2-7$, demonstrating better-constrained model surfaces in this region.
The distribution of best-fit parameters $x_1$ and $c$ of SNe~Ia in the whole training sample and new \textit{HST}/STIS sample are shown in Figure~\ref{fig:paramdist};
the average SALT3 parameters of the new \textit{HST} UV sample are $\bar x_1 = -0.05$ and $\bar c = 0.048$ with standard deviation $\sigma_{x_1} = 0.91$ and $\sigma_{c} = 0.07$, marginally redder than the average yet still consistent with the expected distribution for SN~Ia sample overall. 

Figure \ref{fig:photooverplot} reveals a smaller amplitude of variation in $M_1$ and less scatter near peak in the $u$ band, which may also be due to having better-defined model surfaces in the UV.  The other surfaces are broadly consistent with the previous SALT3 model, albeit with some deviation in $RI$ bands potentially due to the shift in the definition of the $M_0$ and $M_1$ surfaces, changes in the color law that propagate to the $M_0$/$M_1$ surfaces by changing the best-fit $c$ parameters, and the influence of regularization factors. 
Meanwhile, in Figure~\ref{fig:photooverplot}, the $M_0$ component in SDSS-$u$ shows a dip in the first day after explosion. Such phenomena are expected due to the uncertainties in photometric measurements, especially during the pre-explosion and faint early stage.  This could indicate variations in explosion time or insufficient data constraints on the cubic spline interpolation at early times.

Figure~\ref{fig:SALTmodelcomp} shows the comparison between the $M_0$ and $M_1$ components of the SALT3-K21 and SALT3-UV models. 
Due to the scarcity and limited S/N of the UV spectroscopic data and the overall stricter regularization restrictions, both $M_0$ and $M_1$ components of the previous SALT3-K21 model flatten at the blue end with extremely large uncertainties, showing a featureless UV continuum that does not vanish significantly at the shorter wavelength limit as expected.
In comparison, the SALT3-UV model has significantly lower uncertainties and can successfully reproduce the blended line features below $3200$\AA\ from Fe\,\three, Co\,\three\ and Mg\,\two, as well as a vanishing UV continuum at the short wavelength limit.  In particular we see a $>3$x  improvement in the uncertainty level of $M_0$ and $M_1$ below $\sim2800$\AA\ in all phases. Furthermore, a comparison of the best-fit SALT3-K21 and SALT3-UV models with a few selected SNe~Ia UV spectra in the new training sample are shown in Figure~\ref{fig:speccomparison}, highlighting the dramatic improvements in reproducing crowded line features as well as matching the continuum level.

\subsection{Caveats in the SALT3-UV Model}\label{sec:UVissue}

Despite the improvements in model performance, SALT3-UV model is still subject to certain limitations arising from the available data and current training scheme, as detailed in the following discussion.
As shown in Figure~\ref{fig:data-density}, the UV spectroscopic training sample only has sufficient coverage between $-10$ to $+20$ days relative to the B-band maximum. Outside this phase range, the coverage is extremely sparse. Before $-10$ days, the sample only contains 2 and 3 spectra from SN~2011fe and SN~2022hrs, respectively, and after $+20$ days there are only 3 spectra from SN~2011fe and 1 spectrum from SN~2013dy. In particular, we would like to caution that SN~2011fe has been shown to have excess flux in UV \citep{foley16}. 
Thus, our model could be unreliable and biased outside this phase range, and we recommend against its use. 
However, this limitation only has a minimal impact as few SNe~Ia are observable in these phases in rest-frame UV.

Apart from that, we find that the UV model is most reliable when we adopt relatively small regularization factors in our model training to better reproduce the sharp spectral and temporal variations observed at shorter wavelengths. However, this also results in insufficiently smooth model surfaces and thus less reliable distance estimates at longer wavelength.
Combining these factors together, we restrict our subsequent fitting and cosmological analysis to phases between $-10$ to $+20$ days throughout this paper.

Additionally, our UV SN~Ia sample unavoidably includes ZTF and ATLAS photometry that are not yet calibrated to cosmological precision as mentioned in Section~\ref{sec:photo}. The uncertainties and zero-point offsets of these data may subtly influence the model surfaces in our training. Nonetheless, with the on-going recalibration of ZTF and ATLAS data \citep{Lacroix25,Marlin25}, this issue will be resolved by re-training with updated photometry in the near future.

\begin{figure*}
\centering
\includegraphics[width=0.99\linewidth]{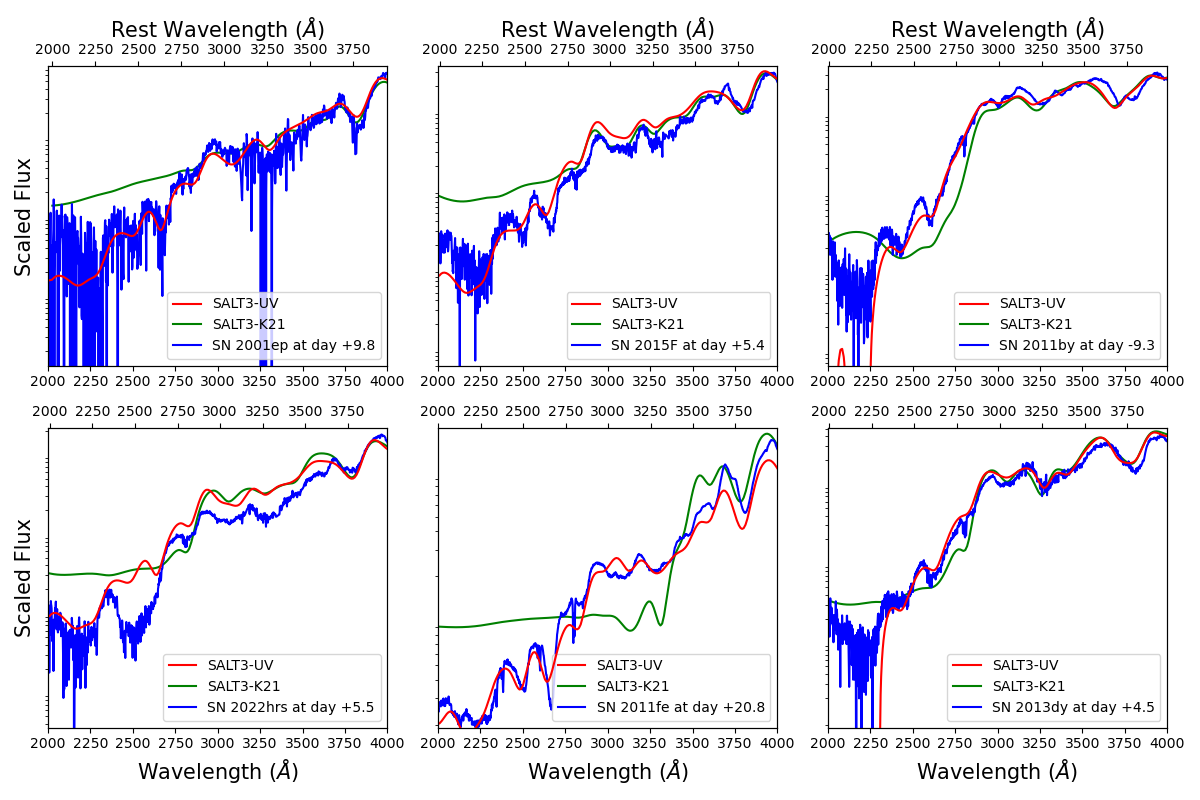}
    \caption{The comparison between the UV spectra (blue) of SNe~Ia in the training sample with the best-fit SALT3-K21 (green) and SALT3-UV (red) models in logarithmic scale. The models are photometrically fitted to individual SNe.
    }
    \label{fig:speccomparison}
\end{figure*}

Lastly, Figure~\ref{fig:speccomparison} also reveals the emergence of extremely low or negative flux and a corresponding reduction in model performance below $\sim 2300$\AA\ for some SNe in our sample. These phenomena occur for SNe~Ia with $x_1 >0$ due to the vanishing $M_0$ component and negative $M_1$ component at shortest wavelength. This is likely attributable to numerical artifacts stemming from noise when instrumental measurements fall below or near the physical detection limit, or the smoothing effect due to the regularization when the flux abruptly decreases at the shortest wavelengths. As can be seen from Figure~\ref{fig:SALTmodelcomp}, those numerical anomalies are well characterized within the uncertainty range of the $M_0$ and $M_1$ components. Furthermore, as upcoming surveys rarely cover $\lambda_\textrm{rest}\lesssim2300$\AA, this caveat should not be considered a critical deficiency of the current model. In preparation for potential rest-frame far-UV surveys in next decade, however, improvements in the training sample are still necessary to optimize the model performance at the shortest wavelength range.

\subsection{Training High- and Low-Redshift UV Models}\label{sec:ztraining}

As noted earlier, several factors could result in redshift evolution of the UV properties of SNe~Ia, including metallicity effects \citep{Pan20}, variations in progenitor mass \citep{Polin19}, and possible changes in the relative contribution of different explosion channels, if they exist, over cosmic time \citep{Rigault13}. Such dependencies could introduce systematic differences in rest-frame UV observables and must be considered when using UV data for cosmological distance measurements.

To test for such evolution, and its impact on cosmology (Section \ref{sec:validation}), in addition to the SALT3-UV model trained with the full sample, we train low- and high-redshift UV models separately.  The low-$z$ model is trained on a subsample only containing $z < 0.1$ SNe\,Ia to exclude the influence of the high-$z$ rest-frame UV data. The UV spectroscopic data are dominated by the {\it HST} spectra from the new sample introduced in this work. There are a total of 351 SNe~Ia and 87 UV spectra from 30 total SNe in the low-$z$ training sample.

The training of the high-$z$ SALT3 UV model includes all of the SNe~Ia used in the SALT3-UV training sample but excludes all spectroscopic data below 3400\AA\ in the rest-frame for SNe~Ia below $z<0.1$ as well as all observer-frame $u/U$-band data. 
The high-$z$ subsample contains 42 spectra extending to the rest-frame UV from 36 SNe~Ia, which have a median $z = 0.533$. The high-$z$ rest-frame UV data are dominated by photometry from the Pan-STARRS, DES, and SNLS samples and spectra from the SNLS sample.  However,
in order to maintain consistent optical properties and avoid the fine-tuning of the regularization that would be necessary with a greatly reduced sample size, the rest of the low-$z$ optical/NIR are also included in the training sample, which allows us to evaluate the difference in the rest-frame UV evolution and its effect on cosmology alone.
 The photometric and spectroscopic data density of each sample are shown in Figure~\ref{fig:nspec-z} and the left and middle panels of Figure~\ref{fig:data-density}. 
Noticeably, the rest-frame UV spectra in the high-$z$ SNLS sample are primarily distributed between $-10$ to $+5$ days relative to peak, likely due to the need to observe high-$z$ SNe\,Ia near their peak for sufficient S/N.
Thus, the high-$z$ SALT3 UV model is barely constrained by spectroscopic data outside this phase range, and the comparisons in the following sections are restricted in this time range.

\subsection{Comparison between High- and Low-$z$ UV Models }\label{sec:zcomp}

\begin{figure*}
    \centering
    \includegraphics[trim={0 1cm 0 1cm}, clip,width=0.49\linewidth]{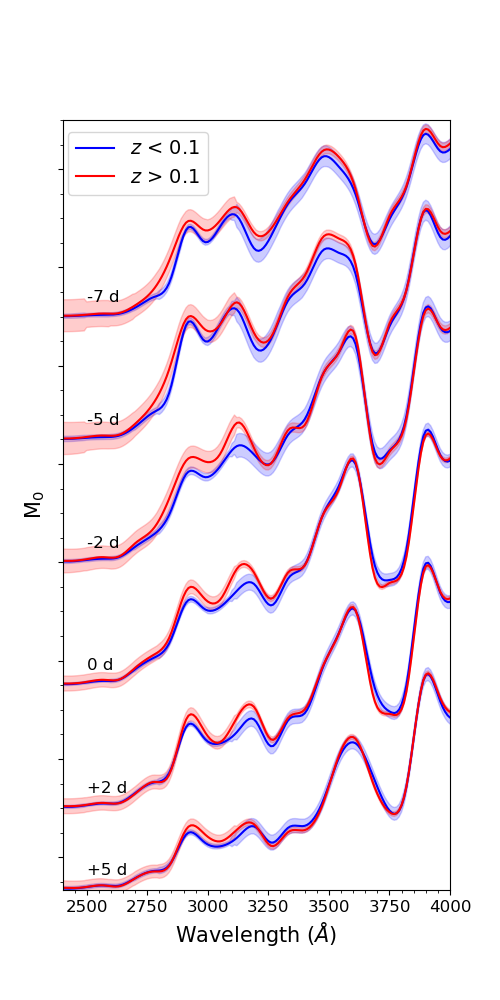}
    \includegraphics[trim={0 1cm 0 1cm}, clip,width=0.49\linewidth]{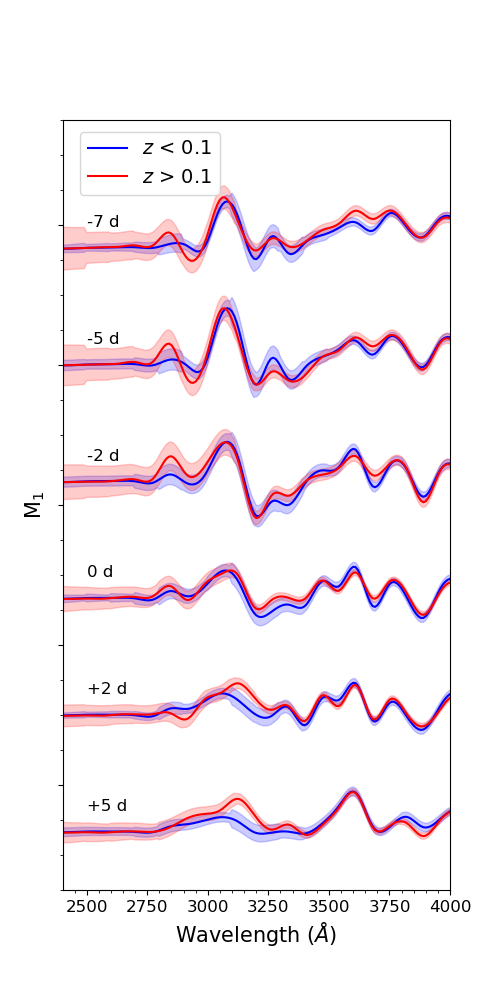}
    \caption{Comparison of the near-peak $M_0/M_1$ components of the SALT3-UV model trained with the low-$z$ ($z<0.1$, blue) and high-$z$ ($z>0.1$, red) samples. The shadowed regions indicate the error range. 
    }
    \label{fig:modelcomp}
\end{figure*}

Our results show some subtle but discernible divergence between the low- and high-$z$ SALT3-UV models.
Figure \ref{fig:modelcomp} shows the comparison between the $M_0$ and $M_1$ components of the resulting low- and high-$z$ models. We see hints of differences in the spectra blueward of $\sim$3200\AA. In particular, the line features in the $M_0$ component of the high-$z$ model are significantly stronger near maximum light. The high-$z$ model also shows stronger line components in $M_1$ especially at post-peak phase, possibly indicating a larger intrinsic variation across different SNe~Ia. 
The high-$z$ model also shows a stronger UV continuum in $M_0$ below $\sim$3200\AA. Overall, the relative difference in flux grows larger at shorter wavelengths, though the high-$z$ model suffers from the lack of data at the short wavelength regime, reflected in the significantly larger uncertainties.

A clear distinction can also be seen in the color law, where the low- and high-$z$ models have more significant differences in the UV, with $\Delta CL(\lambda)\sim 0.1-0.15$ for $2000\text{\AA} <\lambda<3000\text{\AA}$. We can quantify the effect of the color-law difference for a typical SN~Ia sample by $\sigma_c\cdot \Delta CL(\lambda)$, where $\sigma_c = 0.1$ is the standard deviation of the distribution of the color law parameter $c$. As shown in the right panel of Fig.~\ref{fig:colorscatter}, the difference in the color-law alone can result in a difference of $\sim 0.01$~mag in rest-frame UV. It is also worth noting that the change in the UV part of the color-law could induce correlated variation in the optical; if including the UV improves the precision of best-fit color-law parameters, it could propagate into the optical regime and subtly affect the color-law at longer wavelengths.  

\movetabledown=1.5in
\begin{rotatetable*}
\centering
\begin{deluxetable*}{ccccccccccc}
\tablecaption{\label{tab:sninfo} Summary of SNe~Ia with HST/STIS G230L/G430L/G750L data in this work. }
\tablehead{\colhead{SN name} & 
\colhead{$z_{\rm Helio}$} & 
\colhead{Peak MJD} & 
\colhead{$x_0$} & 
\colhead{$x_1$} & 
\colhead{$c$} & 
\colhead{$N_{\rm spec}$} & 
\colhead{$N_{\rm spec}^{\rm HST/STIS}$} & 
\colhead{Photometry surveys} & 
\colhead{$E(B-V)_{\rm MW}$} & 
\colhead{HST Program}}
\startdata
SN~2001eh & 0.037129 & 52170.051 & $0.0041$& $1.7055$& $-0.0440$ &2 &2 & CFA3$^{1}$, LOSS$^{2}$ &0.062 & GO-9114$^{3,4}$ \\
SN~2001ep & 0.01301 & 52200.075 & $0.0205$& $-1.3712$& $0.0641$ & 23 & 4 & CFA3$^{1}$ & 0.041 & GO-9114$^{3,4}$\\
SN~2011by& 0.00284 & 55690.918 & $0.1259$& $-0.2754$& $0.0032$ & 2 & 2 & LOSS$^{5, 6}$& 0.012 & GO-12298$^{7,8}$\\
SN~2011fe& 0.000823 & 55814.5 & $1.9402$& $-0.9162$& $-0.0265$ & 5 & 5 & \textit{Swift}$^{9, 10}$, LOSS$^{6}$ & 0.0075 & GO-12298$^{7,8,11,12}$\\
SN~2013dy& 0.003939 & 56501.1 & $0.1521$& $0.9308$& $0.0582$ & 10 & 10 & \textit{Swift}$^{9, 10}$, LOSS$^{6}$& 0.132 & GO-13286$^{13}$\\
ASASSN-14lp& 0.0034 & 57015.888 & $0.2574$& $0.4008$& $0.1820$ & 10 & 10 & CSP$^{14}$ & 0.0213 & GO-13646$^{11, 15}$ \\
SN~2015F& 0.00489 & 57106.5 & $0.1441$& $-1.1280$& $0.0566$ & 10 & 10 & \textit{Swift}$^{9, 10}$, CSP$^{14}$& 0.1749 & GO-13646$^{11}$\\
SN~2017cbv$^*$& 0.004113 & 57826.1 & $-$& $-$& $-$ & 1 & 1 & CSP$^{14}$& 0.1453 & GO-14925$^{16}$\\
SN~2017erp& 0.006261 & 57937.3 & $0.0834$& $0.1818$& $0.0775$ & 4 & 4 & LOSS$^{6}$ & 0.093 & GO-14665$^{17}$\\
SN~2020uxz& 0.008246 & 59143.362 & $0.0719$& $0.4710$& $-0.0161$ & 26 & 1 & ZTF$^{18,19}$, PS1$^{20}$ & 0.033 & GO-16238$^{21}$\\
SN~2020yvu& 0.01 & 59174.438 & $0.0175$& $0.8767$& $-0.0690$ & 1 & 1 & ZTF$^{18,19}$, ATLAS$^{22}$ & 0.096 & GO-16238$^{21}$\\
SN~2021J& 0.002388 & 59232.7294 & $0.1796$& $-1.0677$& $0.10619$ &20 &2 & ZTF$^{18}$, PS1$^{20}$, ATLAS$^{22}$ & 0.017 & GO-16238$^{21}$ \\
SN~2021dov& 0.0127 & 59288.31 & $0.0211$& $1.0792$& $0.0549$ & 1 & 1 & ZTF$^{18}$, PS1$^{20}$ & 0.03 & GO-16238$^{21}$\\
SN~2021fxy& 0.009483 & 59307.4 & $0.0579$& $0.3573$& $0.0124$ & 4 & 4 & ZTF$^{18}$, ATLAS$^{22}$ & 0.08195 & GO-16221$^{23}$\\
SN~2021hiz& 0.0033 & 59320.310 & $0.0923$& $-0.6947$& $0.0664$ & 1 & 1 & ZTF$^{18}$, PS1$^{20}$ & 0.022 & GO-16238$^{21}$\\
SN~2022hrs& 0.0047 & 59700.520 & $0.1689$& $-1.2564$& $0.2000$ &7 &7 & ZTF$^{18}$, ATLAS$^{22}$&0.023 & GO-16190$^{24}$, 16690$^{25}$\\
SN~2023bee& 0.0067 & 60002.0 & $0.1054$& $1.5544$& $-1.153$ & 1 & 1 & PS1$^{26}$ & 0.014 & GO-17170$^{27}$\\
SN~2023gft& 0.0098 & 60069.757 & $0.0403$& $0.0246$& $0.0305$ & 1 &1 & ZTF$^{18}$, ATLAS$^{22}$ & 0.053 & GO-17170$^{27}$\\
\enddata
\tablecomments{
$^{*}$ The light curve fit to SN~2017cbv fails to converge in training process and thus was excluded.\\
$^{1}$ \cite{CfA3_Hicken09}; 
$^{2}$ \cite{Ganeshalingam10}; 
$^{3}$ \cite{Sauer08}; 
$^{4}$ \cite{Foley08}; 
$^{5}$ \cite{Silverman13}; 
$^{6}$ \cite{Stahl19}; 
$^{7}$ \cite{foley13}; 
$^{8}$ \cite{Graham15}; 
$^{9}$ \cite{Brown14}; 
$^{10}$ \href{https://pbrown801.github.io/SOUSA/}{https://pbrown801.github.io/SOUSA/}; 
$^{11}$ \cite{foley16}; 
$^{12}$ \cite{Mazzali14}; 
$^{13}$ \cite{Pan15}; 
$^{14}$ \cite{Krisciunas17}; 
$^{15}$ \cite{Foley13b}; 
$^{16}$ \cite{2017hst..prop14925F}; 
$^{17}$ \cite{Brown19}; 
$^{18}$ \cite{ZTF_Masci19}; 
$^{19}$ \cite{Rigault25}; 
$^{20}$ \cite{Aleo23}; 
$^{21}$ \cite{2020hst..prop16238F}; 
$^{22}$ \cite{Tonry18}; 
$^{23}$ \cite{DerKacy23}; 
$^{24}$ \cite{HSTGO16190}; 
$^{25}$ \cite{2021hst..prop16690F}; 
$^{26}$ \cite{23bee}; 
$^{27}$ \cite{HSTGO17170}
}
\end{deluxetable*}
\end{rotatetable*}

\begin{figure*}
    \centering
    \includegraphics[width=0.48\linewidth]{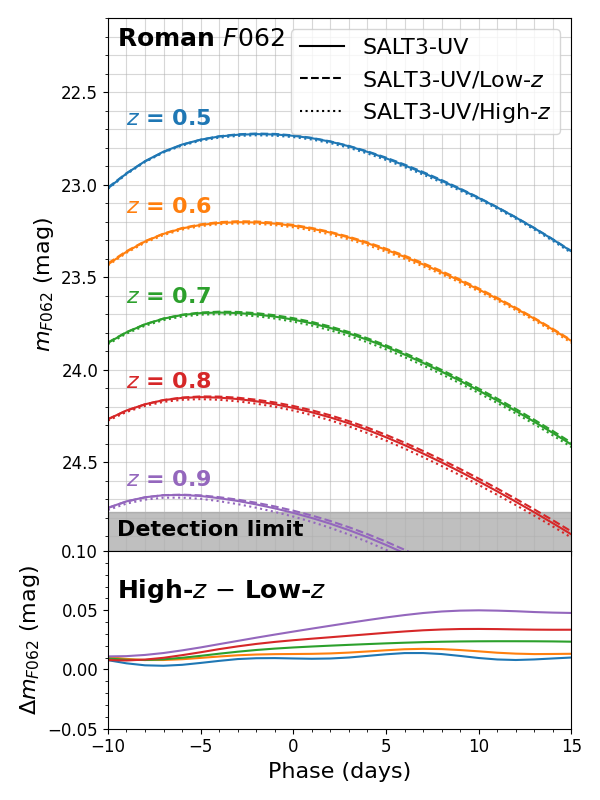}
    \includegraphics[width=0.48\linewidth]{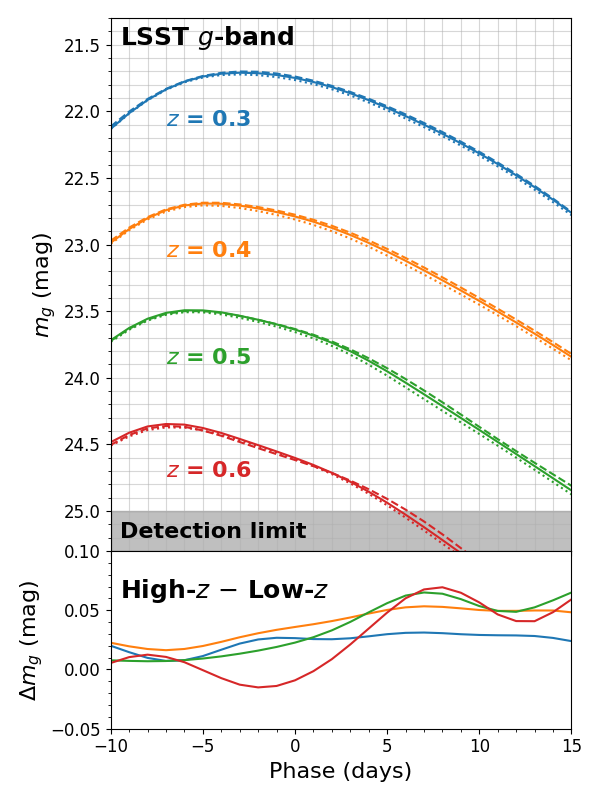}
    \includegraphics[width=0.48\linewidth]{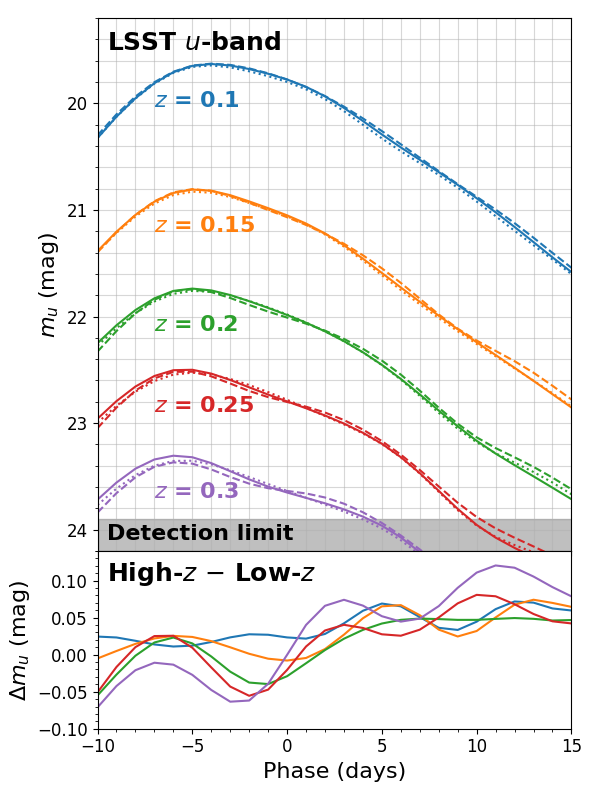}
    \includegraphics[width=0.48\linewidth]{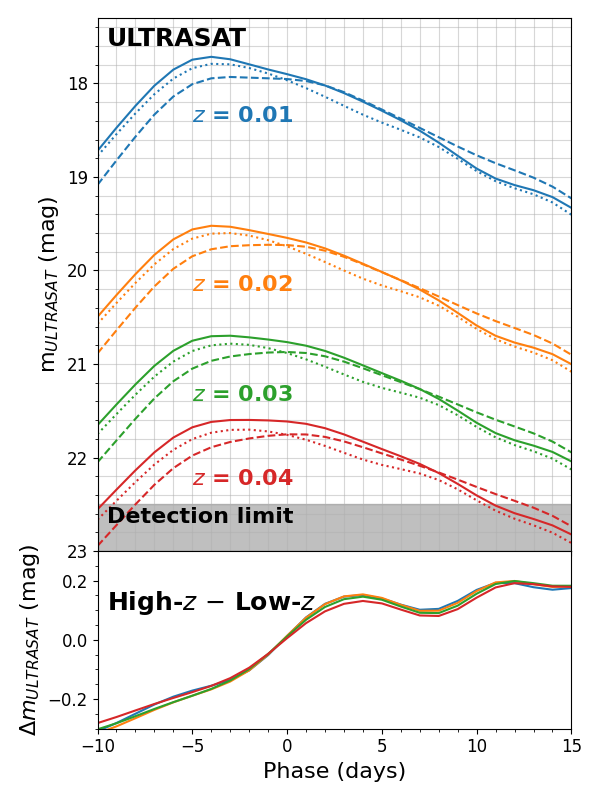}
 
    \caption{Comparison of the mean light curves of the SALT3-UV, low- and high-$z$ models in different bands across different redshift ranges. The phase is relative to the $B$-band maximum, and the grey regions represent the detection limit of corresponding telescope filters \citep{Rose21, lsst19,ULTRASAT}. Note that the high-$z$ dataset lacks rest-frame UV spectroscopic data below $-10$ days and above +5 days --- the models should not be trusted in that range --- and the flattening of the high-$z$ model in the ULTRASAT band in this phase range is potentially a result of stronger regularization due to the low data density.}
    \label{fig:lccomparison}
\end{figure*}

In Fig.~\ref{fig:lccomparison}, for each model we present the `average' light curve in the ULTRASAT, the LSST $u$ and $g$, and the {\it Roman F062} bands, spanning from low redshift to their respective detection limits. 
We note that the high redshift SN~Ia sample is influenced by the Malmquist bias and thus has larger $\bar x_1$. While such selection effects shift the model surface due to the requirement that $\bar x_1$ and $\bar c$ have a mean of zero in each model training, these definitional shifts do not affect cosmological parameter measurements as it has been taken into account in standardization and cosmological analysis \citep{Rubin16}. That said, to compare the `average' model at the global mean, we need to correct for this selection bias in the model surface caused by the Malmquist bias. Therefore, we corrected for the difference in $\bar x_1$ and $\bar c$ of the low- versus high-$z$ subsamples and use that as the `average' light curves instead of $M_0$ component.
Again, we caution that the high-$z$ model is unreliable beyond +5 days as the training sample lacks rest-frame UV spectra at later phases.

In the $u$-band, the difference between the low- and high-$z$ models at $z>0.2$ can be up to $\gtrsim0.05$ mag before $B$-band maximum. The high-$z$ model also shows a faster decline in magnitude after peak. The deviation in the LSST $g$ and Roman {\it F062} bands are around $\sim0.05$ and $\sim 0.03$~mag when approaching their detection limit at redshift around $z\gtrsim0.5$ and $0.8$ respectively. The implications of such deviations on cosmological parameter measurements will be discussed in Section~\ref{sec:validation}. 

As the next generation UV photometric survey, Ultraviolet Transient Astronomy Satellite (ULTRASAT) will conduct a wide-field time domain survey with both low-cadence ($\sim1-4$ days) and high-cadence ($300$ seconds) modes and will observe $\gtrsim200$ SNe~Ia \citep{ULTRASAT}. ULTRASAT's bandpass will cover $\lambda\sim 2300-2900$\AA\ with $\lambda_\textrm{eff} \sim 2600$\AA\ , and Figure~\ref{fig:lccomparison} shows a $\sim 0.2-0.3$ mag difference between the high-$z$ and SALT3-UV/low-$z$ models in ULTRASAT. Moreover, the high-$z$ model exhibits a faster rise compared to the low-$z$ model, reaching a peak in the ULTRASAT band $\sim 3-5$ days earlier as well as a faster decline afterwards. Nonetheless, the high-$z$ model remains poorly constrained at these wavelengths regime due to the scarcity and limited S/N of relevant data, making it partially susceptible to regularization effect. Additional high-$z$ data is necessary to confirm the observed divergence at the short-wavelength limit.

\section{Distances and Cosmology Constraints} \label{sec:validation}

In this section we use the three SALT3-UV models trained in this work to quantify how these different models affect SN\,Ia distance measurements and whether the redshift dependence in the UV could have significant impact on SN\,Ia cosmological parameter constraints. 

\subsection{Generating Simulated Training Samples}

Following previous work from \citet{kenworthy_salt3_2021} and \citet{Pierel_salt3nir_2022}, we use the \snana\  software \citep{snana} to generate realistic simulations of our training data. \snana\  is a suite of simulation and light-curve fitting tools that can simulate SN\,Ia data based on real observing conditions, SN\,Ia models, intrinsic distributions of SN parameters such as $x_1$ and $c$, and known survey selection effects \citep{snana}.

In this work, we generate a simplified SN\,Ia dataset by simulating the Foundation \citep{Foley18, Jones19}, PS1 \citep{Scolnic18}, DES \citep{Kessler19, Abbott19} and SNLS \citep{Astier06} SN\,Ia samples. Those samples combined together have sufficient coverage from low to high redshift up to $z \sim 0.8$.  
The choices of simulation parameters follow \citet{Scolnic19}, \citet{Kessler19} and \citet{Jones19} and we adopt the method of \citet{kenworthy_salt3_2021} of simulating SNe that have the same cadence, redshift, and best-fit $c$ and $x_1$ of each light curve to ensure that we are reproducing the observed sample demographics while using our trained SALT3-UV model as the basis for our simulation. We simulate 20 realizations of the Foundation and SNLS samples as the low- and high$-z$ samples, and another 20 realizations of the DES and PS1 samples to fill in the intermediate redshift range for complete redshift coverage. All the simulations are based on the SALT3-UV model we trained. We further bootstrap the combined sample to create 25 resamples for estimating the in-sample uncertainty of cosmological parameters in the following analysis.  We omit bias corrections in this analysis as we report only on the differences in $w$ resulting from different models, rather than the measurement of $w$ itself; this implicitly assumes that the bias correction does not significantly vary between models, which is a reasonable assumption given that the optical magnitudes used to discover SNe are nearly identical between models.

\subsection{Distance Measurements with Simulated Samples }

We again use the light-curve fitting and distance modulus calculation functions in \snana\ to fit the simulated SN~Ia samples with our trained low- and high-$z$ models as described below. \snana's light-curve fits return the $x_0$, $x_1$, and $c$ parameters, and their associated variances/covariances, which can then be used to measure distances via the Tripp equation:

\begin{equation}
    \mu = m_B + \alpha \cdot x_1 - \beta \cdot c -\mathcal{M},
    \label{eqn2}
\end{equation}

\noindent where $m_B$ is 
the log of the amplitude $x_0$ plus a constant, $\mathcal{M}$ is the absolute peak magnitude assuming some nominal value of the Hubble constant, H$_0$. The $\alpha$ and $\beta$ parameters are nuisance parameters which we measure for our samples.

The simulated samples are first fit with the low- and high-$z$ SALT3-UV models using \snana's {\tt snlc\_fit} light curve fitting module. 
To determine whether the redshift-dependent difference in the UV model causes bias in $w$ and evaluate how it will influence future cosmological surveys, we fit the model with $griz$ and $riz$-band data separately as the former includes the rest-frame UV data at high-$z$ ($g-$band at $z\gtrsim0.3$) while the latter does not.
Similar to other cosmological analyses, we apply standard selection cuts of $-3 < x_1 < 3$, $-0.3 < c < 0.3$, $\sigma_{x_1} < 1$ for further analysis \citep[e.g.,][]{Brout22}. As different models will result in slightly different data being cut in fitting, only the common SNe~Ia passing data cut for both models are used in later cosmology fitting.

To measure the nuissance parameters $\alpha$ and $\beta$ and ultimately $\mu$, we use \snana 's {\tt SALT2mu} program, which follows the method of \citet{Marriner11} to estimate a series of binned distances for our sample while simultaneously measuring the nuisance parameters. Fitting binned distances and nuisance parameters together avoids the possibility of biased measurements due to non-$\Lambda$CDM cosmology or uncorrected Malmquist bias. 

Figure~\ref{fig:HubbleDiagram} shows the difference in the measured distance modulus $\mu$ calculated by Eqn.~\ref{eqn2} between low- and high-$z$ models as a function of redshift for one of the resamples, corrected by the mean offset between the two sets of distance moduli (such offsets are marginalized over in cosmological analyses and do not influence the final results). 
In general, when fitting with $riz$-bands, the measured $\mu$ from the low- and high-$z$ models are consistent, while a noticeable trend in the difference between the two sets of distance moduli ($\Delta\mu$) appears as a function of redshift when the $g$-band is included; the mean $\Delta\mu$ is up to $\sim 0.02-0.03$ mag at maximum. 
Such redshift dependence is concerning, as it could cause a bias in cosmological parameter measurements, in particular in the dark energy equation-of-state parameter $w$.
Note that the difference at low and intermediate redshift in the $griz$ fit is a result of the aforementioned correction in $\bar \mu$ and differences in nuisance parameters caused by high-$z$ measurements;  this difference does not necessarily indicate changes in the rest-frame optical models or other model differences affecting low-$z$ SNe.

\subsection{Systematic Bias in Cosmological Parameter Measurements with Rest-frame UV Photometry}

We next use \snana 's {\tt wfit} module to estimate cosmological parameters after applying a prior on the cosmic matter density $\Omega_m$ from the WMAP 2009 cosmology results \citep{WMAP09}. 
We do not attempt a complete cosmological analysis, but instead wish to demonstrate the relative difference in $w$ measurements introduced by different UV models under this simplified simulation setting.
The actual measured value of $w$ does not have real physical implications as we do not correct for Malmquist bias as noted previously. Instead, we calculate the difference in $\Delta w$ between low- and high-$z$ models.  We use 25 bootstrapped resamples to quantify the systematic bias introduced by different choices of the UV model.

\begin{figure}
    \centering
    \includegraphics[width=\linewidth]{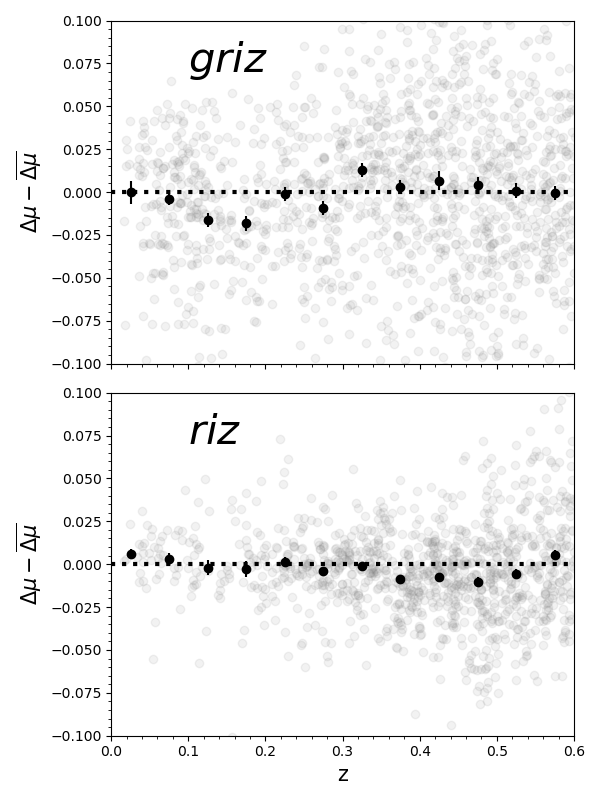}
    \caption{The difference in the distance modulus of one simulated SN~Ia resample fitted by the low- and high-$z$ SALT3 UV models ($\Delta\mu$).  Black dots denote the mean $\Delta\mu$ and uncertainties in each redshift bin. The plot has been corrected by the mean offset in distance modulus between fits using each SALT3 UV model ($\overline{\Delta\mu}$).
    }
    \label{fig:HubbleDiagram}
\end{figure}

Due to the redshift dependence in $\Delta\mu$ seen in Fig.~\ref{fig:HubbleDiagram}, when fitting with the $griz$ bands we see a mean $\Delta w$ of $0.0216\pm0.0030$ among 25 resamples, while $\Delta w$ is marginally significant at $0.0092\pm0.0052$ for the $riz$-only fit. We caution that the presented uncertainty budget in $\Delta w$ only reflects the in-sample statistical uncertainties and does not include the propagated uncertainties in model surfaces.
Still, this deviation is a strong indication of the potential systematic bias in high-$z$ cosmological measurements due to short-wavelength data.
Given that the only difference between the two models is whether the low-$z$ UV data are included or not, we conclude that the most significant deviations come from the difference in the rest-frame UV model surfaces and color laws between the low- and high-$z$ models.

\section{Discussion} \label{sec:discussion}

\subsection{Constraints on the SN~Ia redshift evolution}

Recent studies have explored the potential redshift evolution of SNe~Ia with high-$z$ observations from JWST in the optical and NIR \citep{Pierel24b, Pierel25}, and complementary rest-frame UV studies will be crucial in tracing this potential evolution. 
As revealed by previous studies, the UV properties of SNe~Ia are more diverse and extremely sensitive to their progenitor properties and explosion mechanisms \citep{foley16, Milne2015}. 
If these factors, e.g. metallicity, binary mass, or the relative ratio of different progenitor channels, evolve with redshift, a systematic bias may arise in cosmological distance measurements when including the rest-frame UV data of SNe~Ia \citep{Milne2015}. 
The divergence between UV SED templates of SNe~Ia at low- and high-$z$ observed in our results is consistent with these results.
In a comparison study between low-$z$ and intermediate-$z$ SNe~Ia with a limited sample, \cite{Foley12a} finds that although the optical spectra of low- and intermediate-z ($0.11\lesssim z \lesssim0.37$) SNe~Ia are largely consistent, the intermediate-$z$ SNe~Ia exhibit a $\gtrsim20\%$ increase in rest-frame UV flux. \cite{Foley12a} also discovered that the relationship between the peak luminosity and flux ratio at 2770 and 2900\AA\ is different between the low- and intermediate-$z$ sample. 
These observed trends agree with the $\sim0.1-0.3$ mag increase in the $u$ and ULTRASAT magnitudes seen in our models, as well as the changes in line intensity in high-$z$ SNe~Ia templates compared to low-$z$. 

One leading explanations of the divergence between low- and high-$z$ SALT3-UV model could be the difference in progenitor metallicity and/or host metallicity. Previous studies at low-$z$ have revealed that SNe~Ia in galaxies with lower metallicity tend to have stronger UV flux below $2700$\AA\ \citep{Pan20}. As revealed in Section~\ref{sec:zcomp}, the high$-z$ SALT3-UV model tends to be more UV-bright than the low-$z$ model. This observed trend qualitatively matches cosmic metallicity evolution, making it a plausible explanation.
However, it is unclear whether the metallicity evolution is the dominant factor due to the difficulties in obtaining a solid metallicity estimation of the host galaxies at high-$z$. Another speculation could be the evolution of progenitor channels. \cite{Rigault13} noted that SNe~Ia with local H$\alpha$ emission, indicative of young star population, tend to have redder color. They further propose that such trend could potentially be explained by two separate population of SNe~Ia, with one broadly exists in all environment while the other occurs exclusively in passive environment. 
If multiple progenitor channels, e.g. single- and double-degenerate systems, contribute to the normal SNe~Ia population simultaneously, the evolution of relative rate between channels with cosmic time could result in redshift dependency. A larger UV sample at both low- and high-$z$ across diverse environment is necessary to disentangle the puzzle of the UV divergence.

\subsection{Implications for Cosmology}

The comparison between the low- and high-$z$ models reveal some discernible differences. 
Specifically, we see potential systematics that could affect the LSST $u-$band at $z>0.1$, $g-$band at $z\gtrsim0.3$, and {\it Roman F062} band at $z\gtrsim0.6$.
Such differences, if correct, imply that the rest-frame UV data have the potential to introduce nonnegligible systematic bias in future cosmological measurements. 
In particular, our simulations demonstrate that such divergence could potentially introduce a $\sim 0.02$ bias in $w$ if high-$z$ $g$-band data are included, which illustrates the sensitivity of dark energy measurements to our current uncertainties in the redshift evolution of SN\,Ia UV SEDs.

More spectroscopic observations from {\it HST} and data from future UV missions such as ULTRASAT and {\it UVEX} will help further pin down the UV SED model of SNe~Ia and confirm the differences we observe in low- and high-$z$ UV models. 
For cosmological usage, in the future it will be beneficial to determine if this potential redshift evolution can be calibrated by other parameters with a similar evolutionary trend, e.g., host-galaxy metallicity. As correlated $z$-dependence may also occur in the optical or infrared, though likely to a lesser degree \citep{Pierel24b, Pierel25}, incorporating such calibrations could also be valuable for SN\,Ia studies at other wavelengths.
On the other hand, we note that the Malmquist bias is more obvious for the high-$z$ spectroscopic sample (SNLS); while for the complete high-$z$ sample $\bar x_1=0.0678$ when fitted to the SALT3-K21 model, the subsample with rest-frame UV spectra has a $\bar x_1$ as high as $0.538$. 
Although the SALT model and cosmological analysis are generally robust against the Malmquist bias in spectroscopic sample \citep{Rubin16, Nicolas21, Ruppin25}, it nevertheless underscores the importance of spectroscopic observation of fainter SNe~Ia at high-$z$ to ensure a complete demographic characterization of their UV behaviours.

Currently there exist a number of difficulties in understanding and confirming UV redshift evolution: 1) lack of spectroscopic data in the intermediate redshift bin due to few current missions with UV spectroscopic capability; 2) relatively poor coverage and high uncertainties in host property measurements at redshift $z\gtrsim0.2$  \citep{Qin24}; and 3) the relatively small sample of SNe~Ia with rest-frame UV observations at high S/N, especially at high-$z$. Those gaps awaits deep observations from future optical/UV missions to fill.

\section{Conclusion} \label{sec:conclusion}

Using a newly assembled archival sample of {\it HST}/STIS UV spectra, we have extended the SALT3 model to $2000$\AA, achieving a factor of $\sim3$ reduction in the uncertainties in the $M_0$ and $M_1$ components below $\sim2800$\AA\ and a factor of $\sim 2-7$ improvement in the UV color scatter. 
By incorporating these new UV data and adopting a looser regularization scheme, the updated SALT3-UV model is able to reproduce multiple line features and a smooth continuum that reasonably decreases toward zero at short wavelengths, unlike the previous SALT3-K21 which has an unnaturally flattened continuum below $3000$\AA. Still, the model has caveats due to the limitation of the training sample including: \textbf{1)} phases before $-10$ days or after $+20$ days are not recommended due to the scarcity of the data; \textbf{2)} wavelengths below $2300$\AA\ are less reliable as many spectra in the training sample fall below the detection limit and the regularization function tends to smooth out the template where flux abruptly varies; and \textbf{3)} the model might be marginally affected by the systematic uncertainties in the ATLAS and ZTF calibrations for a few specific SNe in the sample, though this can be mitigated as re-calibrated data become available in the near future. Consequently, it is imperative to be cautious when applying the SALT3-UV model in those marginal cases and cosmological analysis.  Our training sample and \saltshaker\ configuration files are publicly available to facilitate updated trainings when new or recalibrated data become available.

Through training on low- and high-$z$ subsamples, we also find a tentative redshift-dependent difference in the UV model, indicating a potential evolutionary trend across cosmic time. The potential physical causes of such evolution could be the change in the properties of progenitors and/or host galaxies with redshift, such as metallicity. We find potential evolution of up to $\sim0.05, 0.05$ and $0.03$ mag in the LSST $u$, $g$ bands and the {\it Roman F062} band at $z\gtrsim0.2, 0.5$ and $0.8$ respectively. 
Through simulation, we demonstrate that such redshift evolution can introduce a systematic bias of $\sim0.02$ in $w$ if including high-$z$ $g$-band data. This bias, if confirmed, could be a leading source of systematic uncertainty in cosmological analyses \citep[e.g.,][]{Brout22,Abbott24sn}.
Our results therefore suggest that caution must be taken when using rest-frame UV data in future cosmological analyses, e.g. LSST and {\it Roman}, and more work is needed to understand and mitigate potential biases in rest-frame UV bands. Future UV surveys like ULTRASAT ($\gtrsim 200$ SNe Ia per year, private comm) and {\it UVEX} can provide sufficient low-$z$ sample for calibrating the SNe~Ia UV template with host galaxy properties, providing insight on whether the observed redshift dependency trend is intrinsic to SNe~Ia or related to redshift dependent factors. More high-$z$ spectroscopic data, in particular for faint SNe~Ia, would also improve the high-$z$ SALT3-UV model and validate the existence of systematic changes in the SN\,Ia UV SED.\\


The authors would like to acknowledge the constructive discussions with David Rubin, Dillon Brout and Yukei Murakami.

Q.W. is supported by the Sagol Weizmann-MIT Bridge Program and HST grant HST-AR-17024. 
D.O.J. acknowledges support from NSF grants AST-2407632, AST-2429450, and AST-2510993, NASA grant 80NSSC24M0023, and HST/JWST grants HST-GO-17128.028 and JWST-GO-05324.031, awarded by the Space Telescope Science Institute (STScI), which is operated by the Association of Universities for Research in Astronomy, Inc., for NASA, under contract NAS5-26555.
W.D.W. has been enabled by support from the research project grant ‘Understanding the Dynamic Universe’ funded by the Knut and Alice Wallenberg Foundation under Dnr KAW 2018.0067.
Support for M.D. was provided by Schmidt Sciences, LLC.

This work is based on observations made with the NASA/ESA Hubble Space Telescope, obtained from the Mikulski Archive for Space Telescopes (MAST) at the Space Telescope Science Institute, which is operated by the Association of Universities for Research in Astronomy, Inc., under NASA contract NAS5-26555. Support to MAST for these data is provided by the NASA Office of Space Science via grant NAG5–7584 and by other grants and contracts. These observations are associated with program GO-9114, 11721, 12298, 12582, 13286, 13646, 14144, 14925, 14665, 16238, 16690 and 17170. 
Support for program HST-AR-17024 was provided by NASA through a grant from the Space Telescope Science Institute, which is operated by the Association of Universities for Research in Astronomy, Inc., under NASA contract NAS5-26555.
This research used resources of the National Energy Research Scientific Computing Center (NERSC), a Department of Energy User Facility using NERSC award HEP-ERCAP m1727. P.M. acknowledges that this work was partly performed under the auspices of the U.S. Department of Energy by Lawrence Livermore National Laboratory under Contract DE-AC52-07NA27344. The document number is LLNL-JRNL-2013961.


%

\vspace{5mm}
\facilities{HST, ATLAS, Swift, ZTF}


\software{SALTShaker\citep{Guy07, Guy10, Betoule14, kenworthy_salt3_2021}, 
SNANA\citep{snana}, 
astropy \citep{2013A&A...558A..33A,2018AJ....156..123A},
Matplotlib \citep{matplotlib}, 
SciPy \citep[][]{2020SciPy-NMeth}, 
NumPy \citep{2020NumPy-Array}, 
SNCosmo \citep{sncosmo}}


%

\bibliography{sample631}{}

\bibliographystyle{aasjournal}

\end{document}